# The Ni isotopic composition of Ryugu reveals a common accretion region for carbonaceous chondrites

Fridolin Spitzer[*,1], Thorsten Kleine[1], Christoph Burkhardt[1], Timo Hopp[1], Tetsuya Yokoyama[2], Yoshinari Abe[3], Jérôme Aléon[4], Conel M. O'D. Alexander[5], Sachiko Amari[6,7], Yuri Amelin[8], Ken-ichi Bajo[9], Martin Bizzarro[10], Audrey Bouvier[11], Richard W. Carlson[5], Marc Chaussidon[12], Byeon-Gak Choi[13], Nicolas Dauphas[14], Andrew M. Davis[14], Tommaso Di Rocco[15], Wataru Fujiya[16], Ryota Fukai[17], Ikshu Gautam[2], Makiko K. Haba[2], Yuki Hibiya[18,19], Hiroshi Hidaka[20], Hisashi Homma[21], Peter Hoppe[22], Gary R. Huss[23], Kiyohiro Ichida[24], Tsuyoshi Iizuka[25], Trevor R. Ireland[26], Akira Ishikawa[2], Shoichi Itoh[27], Noriyuki Kawasaki[9], Noriko T. Kita[28], Kouki Kitajima[28], Shintaro Komatani[24], Alexander N. Krot[23], Ming-Chang Liu[29], Yuki Masuda[2], Mayu Morita[24], Fréderic Moynier[12], Kazuko Motomura[30], Izumi Nakai[31], Kazuhide Nagashima[23], Ann Nguyen[32], Larry Nittler[33], Morihiko Onose[24], Andreas Pack[15], Changkun Park[34], Laurette Piani[35], Liping Qin[36], Sara S. Russell[37], Naoya Sakamoto[38], Maria Schönbächler[39], Lauren Tafla[29], Haolan Tang[36], Kentaro Terada[40], Yasuko Terada[41], Tomohiro Usui[17], Sohei Wada[9], Meenakshi Wadhwa[33], Richard J. Walker[42], Katsuyuki Yamashita[43], Qing-Zhu Yin[44], Shigekazu Yoneda[45], Edward D. Young[29], Hiroharu Yui[46], Ai-Cheng Zhang[47], Tomoki Nakamura[48], Hiroshi Naraoka[49], Takaaki Noguchi[27], Ryuji Okazaki[49], Kanako Sakamoto[17], Hikaru Yabuta[50], Masanao Abe[17], Akiko Miyazaki[17], Aiko Nakato[17], Masahiro Nishimura[17], Tatsuaki Okada[17], Toru Yada[17], Kasumi Yogata[17], Satoru Nakazawa[17], Takanao Saiki[17], Satoshi Tanaka[17], Fuyuto Terui[51], Yuichi Tsuda[17], Sei-ichiro Watanabe[20], Makoto Yoshikawa[17], Shogo Tachibana[52], Hisayoshi Yurimoto[9]

**Affiliations**

[1]Max Planck Institute for Solar System Research, Justus-von-Liebig-Weg 3, 37077 Göttingen, Germany.

[2]Department of Earth and Planetary Sciences, Tokyo Institute of Technology, Tokyo 152-8551, Japan.

[3]Graduate School of Engineering Materials Science and Engineering, Tokyo Denki University, Tokyo 120-8551, Japan.

[4]Institut de Minéralogie, de Physique des Matériaux et de Cosmochimie, Sorbonne Université, Museum National d'Histoire Naturelle, CNRS UMR 7590, IRD, 75005 Paris, France.

[5]Earth and Planets Laboratory, Carnegie Institution for Science, Washington, DC, 20015, USA.

[6]McDonnell Center for the Space Sciences and Physics Department, Washington University, 35 St. Louis, MO 63130, USA.

[7]Geochemical Research Center, The University of Tokyo, Tokyo, 113-0033, Japan.

[8]Guangzhou Institute of Geochemistry, Chinese Academy of Sciences, Guangzhou, GD 510640, China.

[9]Departement of Natural History Sciences, Hokkaido University, Sapporo 001-0021, Japan.

[10]Centre for Star and Planet Formation, Globe Institute, University of Copenhagen, Copenhagen, K 1350, Denmark.

[11]Bayerisches Geoinstitut, Universität Bayreuth, Bayreuth 95447, Germany.

[12]Université Paris Cité, Institut de Physique du Globe de Paris, CNRS, 75005 Paris, France.
[13]Department of Earth Science Education, Seoul National University, Seoul 08826, Republic of Korea.
[14]Department of the Geophysical Sciences and Enrico Fermi Institute, The University of Chicago, 5734 South Ellis Avenue, Chicago 60637, USA.
[15]Faculty of Geosciences and Geography, University of Göttingen, Göttingen, D-37077, Germany.
[16]Faculty of Science, Ibaraki University, Mito 310-8512, Japan.
[17]ISAS/JSEC, JAXA, Sagamihara 252-5210, Japan.
[18]Research Center for Advanced Science and Technology, The University of Tokyo, Tokyo 153-8904, Japan.
[19]Research Institute for Marine Resources Utilization, Japan Agency for Marine-Earth Science and Technology, Kanagawa, 237-0061, Japan.
[20]Departement of Earth and Planetary Sciences, Nagoya University, Nagoya 464-8601, Japan.
[21]Osaka Application Laboratory, SBUWDX, Rigaku Corporation, Osaka 569-1146, Japan.
[22]Max Planck Institute for Chemistry, Mainz 55128, Germany.
[23]Hawai'i Institute of Geophysics and Planetology, University of Hawai'i at Mānoa; Honolulu, HI 96822, USA.
[24]Analytical Technology, Horiba Techno Service Co., Ltd., Kyoto 601-8125, Japan.
[25]Departement of Earth and Planetary Science, The University of Tokyo, Tokyo 113-0033, Japan.
[26]School of Earth and Environmental Sciences, The University of Queensland, St Lucia QLD 4072, Australia.
[27]Departement of Earth and Planetary Sciences, Kyoto University, Kyoto 606-8502, Japan.
[28]Departement of Geoscience, University of Wisconsin-Madison, Madison, WI 53706, USA.
[29]Departement of Earth, Planetary, and Space Sciences, UCLA, Los Angeles, CA 90095, USA.
[30]Thermal Analysis, Rigaku Corporation, Tokyo 196-8666, Japan.
[31]Department of Applied Chemistry, Tokyo University of Science, Tokyo 162-8601, Japan.
[32]Astromaterials Research and Exploration Science, NASA Johnson Space Center; Houston, TX 77058, USA.
[33]School of Earth and Space Exploration, Arizona State University, Tempe, AZ 85281, USA.
[34]Division of Earth-System Sciences, Korea Polar Research Institute, Incheon 21990, Korea.
[35]Centre de Recherches Pétrographiques et Géochimiques, CNRS - Université de Lorraine, 54500 Nancy, France.
[36]School of Earth and Space Sciences, University of Science and Technology of China, Hefei, Anhui 230026, China.
[37]Department of Earth Sciences, Natural History Museum, London, SW7 5BD, UK.
[38]Isotope Imaging Laboratory, Hokkaido University, Sapporo 001-0021, Japan.
[39]Institute for Geochemistry and Petrology, Department of Earth Sciences, ETH Zurich, Zurich, Switzerland.
[40]Departement of Earth and Space Science, Osaka University, Osaka 560-0043, Japan.
[41]Spectroscopy and Imaging Division, Japan Synchrotron Radiation Research Institute, Hyogo 679-5198 Japan.
[42]Departement of Geology, University of Maryland, College Park, MD 20742, USA.

[43]Graduate School of Natural Science and Technology, Okayama University, Okayama 700-8530, Japan.
[44]Departement of Earth and Planetary Sciences, University of California, Davis, CA 95616, USA.
[45]Departement of Science and Engineering, National Museum of Nature and Science, Tsukuba 305-0005, Japan.
[46]Departement of Chemistry, Tokyo University of Science, Tokyo 162-8601, Japan.
[47]School of Earth Sciences and Engineering, Nanjing University, Nanjing 210023, China.
[48]Department of Earth Science, Tohoku University, Sendai, 980-8578, Japan.
[49]Department of Earth and Planetary Sciences, Kyushu University, Fukuoka 819-0395, Japan.
[50]Earth and Planetary Systems Science Program, Hiroshima University, Higashi-Hiroshima, 739-8526, Japan.
[51]Graduate School of Engineering, Kanagawa Institute of Technology, Atsugi 243-0292, Japan.
[52]UTokyo Organization for Planetary and Space Science, University of Tokyo, Tokyo 113-0033, Japan.

*Corresponding author: spitzer@mps.mpg.de

## Abstract

The isotopic compositions of samples returned from Cb-type asteroid Ryugu and Ivuna-type (CI) chondrites are distinct from other carbonaceous chondrites, which has led to the suggestion that Ryugu/CI chondrites formed in a different region of the accretion disk, possibly around the orbits of Uranus and Neptune. We show that, like for Fe, Ryugu and CI chondrites also have indistinguishable Ni isotope anomalies, which differ from those of other carbonaceous chondrites. We propose that this unique Fe and Ni isotopic composition reflects different accretion efficiencies of small FeNi metal grains among the carbonaceous chondrite parent bodies. The CI chondrites incorporated these grains more efficiently, possibly because they formed at the end of the disk's lifetime, when planetesimal formation was also triggered by photoevaporation of the disk. Isotopic variations among carbonaceous chondrites may thus reflect fractionation of distinct dust components from a common reservoir, implying CI chondrites/Ryugu may have formed in the same region of the accretion disk as other carbonaceous chondrites.

## Introduction

Meteorites are fragments of asteroids and represent material from planetesimal bodies that formed within the first few million years (Ma) of the formation of the solar system. Information about the original formation location of these objects in the solar protoplanetary disk, as well as possible genetic relationships among them, can be inferred from O isotopes (*1*) and nucleosynthetic isotope anomalies in multiple elements (*2–10*). The latter arise from the heterogeneous distribution of presolar materials and vary as a function of formation time and/ or location in the disk (*11–13*). For instance, these isotope anomalies allow for distinguishing between non-carbonaceous (NC) and carbonaceous (CC) type meteorites that represent material from two spatially distinct but co-existing reservoirs, which have been suggested to have been located inside and outside the orbit of Jupiter, respectively (*7*, *14*, *15*).

For most elements, CI (Ivuna-type) chondrites have chemical abundances that are similar to those of the solar photosphere (*16*) and are, therefore, considered the chemically most pristine meteorites. Classified as carbonaceous chondrites, the CI chondrites belong to the CC-type of meteorites, but it has recently been shown that for some elements they have isotopic compositions that are distinct from the compositional cluster defined by all other CC meteorites. This finding is based on the analyses of CI chondrites together with samples of the Cb-type asteroid 162173 Ryugu brought to Earth by the Japan Aerospace Exploration Agency's (JAXA) Hayabusa2 mission. These analyses have not only demonstrated that Ryugu and CI chondrites are mineralogically, chemically, and isotopically similar (*9*, *10*, *17–19*), but also that their nucleosynthetic Fe isotope signature is distinct from all other carbonaceous chondrites (*10*, *20*). Thus, Ryugu and CI chondrites consist of a distinct mix of solar nebula materials than other carbonaceous chondrites. On this basis, it has been suggested that instead of an isotopic dichotomy (NC-CC) there is an isotopic trichotomy (NC-CC-CI), where CI chondrites/Ryugu formed at a greater heliocentric distance than other carbonaceous chondrites, possibly within the Uranus-Neptune region, and were scattered into the terrestrial planet region during growth and migration of the gas giant planets (*10*, *21*). However, the extent to which CI chondrites are isotopically distinct from other carbonaceous chondrites is debated, as is the question of whether materials with isotopic compositions intermediate between CI chondrites and other carbonaceous chondrites exist (*22*, *23*).

Despite the isotopic differences between CI and other carbonaceous chondrites, all carbonaceous chondrites seem to contain some CI chondrite-like material. This is evident from the chemical and isotopic variations among the different groups of carbonaceous chondrites, which, except for Fe isotopes, can readily be accounted for by variable mixtures of the same constituents, namely chondrules, refractory inclusions, and matrix (*24*, *25*). At least some of these components formed in different regions of the solar nebula and as a result have distinct isotopic compositions, and so variations in their abundances lead to isotopic variations among bulk carbonaceous chondrites. On this basis, it has been suggested that the primitive matrix of all carbonaceous chondrites appears to be chemically and isotopically most similar to CI chondrites (*24–27*). Thus, while CI chondrites themselves have been suggested to derive from a different region of the accretion disk than all other carbonaceous chondrites, CI chondrite-like material still appears to be ubiquitously present in the formation regions of all other carbonaceous chondrites. Understanding these seemingly contradictory observations requires determining the processes that led to the distinct isotopic composition of CI chondrites/Ryugu, and how these processes affected the composition of the CI chondrite-like matrix incorporated into carbonaceous chondrites. Resolving these issues is of considerable importance, as it holds clues about the processes affecting solid materials in the disk, the origin of the compositional diversity among planetesimals formed in the outer solar system, and the original formation locations of these objects in the disk.

Here we use Ni isotopes to better constrain the origin of the isotopic differences between Ryugu/CI chondrites and other carbonaceous chondrites. Available Ni isotope data suggest that CI chondrites display the largest $\mu^{62}$Ni and $\mu^{64}$Ni values among carbonaceous chondrites (where $\mu^{i}$Ni is the parts-per-$10^6$ deviation from the terrestrial SRM 986 Ni standard) and are characterized by distinct $\mu^{60}$Ni values compared to all other carbonaceous chondrites (*4*, *28–30*). This makes Ni, which is a siderophile element like Fe and has a similar 50% condensation temperature, uniquely useful for assessing the origin and nature of any isotopic differences between Ryugu/CI chondrites and other carbonaceous chondrites. However, Ni isotope data for CI chondrites are sparse and have only been reported for two samples. Consequently, the Ni isotopic differences between CI chondrites and other carbonaceous chondrites are not well resolved. In addition, no Ryugu samples have yet been analyzed for their Ni isotopic compositions, and therefore it is not known whether CI chondrites and Ryugu share the same Ni isotopic composition. Thus, we have determined the Ni isotopic compositions of four Ryugu samples together with several CI chondrites and other carbonaceous chondrites.

**Results**

The samples investigated in this study include four Ryugu samples (A0106 and A0106-A0107 from the first touchdown site, and C0107 and C0108 from the second touchdown site) together with six carbonaceous chondrites processed alongside the Ryugu samples (*9*), including the CI chondrites Orgueil and Alais, the Mighei-type (CM) Murchison, the Vigarano-type (CV) Allende, and the two ungrouped carbonaceous chondrites Tarda (TD) and Tagish Lake (TL; Materials and Methods). In addition, several other carbonaceous chondrites, including samples from all major carbonaceous chondrite groups (Table 1), were also analyzed. Since Ni can be produced in different stellar sources, any nucleosynthetic Ni isotope variations may reflect anomalies on more than one of the Ni isotopes (*30*), including those used for the correction of natural and instrumental mass fractionation. To identify which Ni isotopes are responsible for the observed variability, it is thus useful to compare the data for different internal normalizations, and like in prior studies (*29*, *31*) we report the Ni isotope data for normalization to either a fixed $^{61}$Ni/$^{58}$Ni or a fixed $^{62}$Ni/$^{61}$Ni. Accordingly, the data are expressed as either $\mu^{60}Ni_{61/58}$, $\mu^{62}Ni_{61/58}$, and $\mu^{64}Ni_{61/58}$ values for the $^{61}$Ni/$^{58}$Ni normalization, or as $\mu^{58}Ni_{62/61}$, $\mu^{60}Ni_{62/61}$, and $\mu^{64}Ni_{62/61}$ values for the $^{62}$Ni/$^{61}$Ni normalization.

The new Ni isotopic data for CM, Ornans-type (CO), CV, and Renazzo-type (CR) chondrites agree well with those reported in previous studies (*4*, *28–30*) and reveal that these chondrites are characterized by negative $\mu^{60}Ni_{61/58}$, and positive $\mu^{62}Ni_{61/58}$ and $\mu^{64}Ni_{61/58}$ values (Fig. 1). The two ungrouped carbonaceous chondrites Tagish Lake (TL) and Tarda (TD), for which no Ni isotopic data have been reported previously, have Ni isotope anomalies that are similar to those of most of the other carbonaceous chondrites. By contrast, all CI chondrites and all four Ryugu samples have larger $\mu^{60}Ni_{61/58}$, $\mu^{62}Ni_{61/58}$, and $\mu^{64}Ni_{61/58}$ values compared to the other carbonaceous chondrites (Fig. 1). As we will show below, given that this difference between CI chondrites/Ryugu and other carbonaceous chondrites is also seen for Fe isotope anomalies (*10*), it can most readily be accounted for by the fractionation of isotopically anomalous FeNi metal grains.

**Table 1. Ni isotopic composition of Ryugu samples and carbonaceous chondrites.** Uncertainties of individual samples represent two standard errors (2 s.e.), where N is the number of measurements. Data is internally normalized to either $^{61}Ni/^{58}Ni$ = 0.016744 or $^{62}Ni/^{61}Ni$ = 3.1884. Ung = ungrouped.

| Sample | Class | N | Norm. $^{61}Ni/^{58}Ni$ | | | Norm. $^{62}Ni/^{61}Ni$ | | |
|---|---|---|---|---|---|---|---|---|
| | | | $\mu^{60}Ni$ | $\mu^{62}Ni$ | $\mu^{64}Ni$ | $\mu^{58}Ni$ | $\mu^{60}Ni$ | $\mu^{64}Ni$ |
| Ryugu samples | | | | | | | | |
| A0106–A0107 | | 24 | -4 ± 3 | 19 ± 6 | 48 ± 10 | 58 ± 18 | 15 ± 9 | -7 ± 12 |
| A0106 | | 20 | -2 ± 4 | 17 ± 8 | 51 ± 14 | 51 ± 24 | 14 ± 11 | 2 ± 15 |
| Ryugu A mean | | | -3 ± 3 | 18 ± 3 | 50 ± 4 | 55 ± 10 | 15 ± 1 | -2 ± 13 |
| C0108 | | 24 | -3 ± 3 | 27 ± 5 | 76 ± 11 | 82 ± 17 | 24 ± 8 | -3 ± 12 |
| C0107 | | 20 | 0 ± 4 | 25 ± 7 | 66 ± 8 | 78 ± 22 | 26 ± 11 | -8 ± 16 |
| Ryugu C mean | | | -1 ± 5 | 26 ± 2 | 71 ± 14 | 80 ± 7 | 25 ± 2 | -6 ± 7 |
| Ryugu mean | | | -2 ± 3 | 22 ± 8 | 60 ± 20 | 67 ± 24 | 20 ± 9 | -4 ± 8 |
| Carbonaceous chondrites | | | | | | | | |
| Orgueil | CI1 | 37 | -3 ± 3 | 15 ± 4 | 43 ± 7 | 46 ± 14 | 12 ± 7 | -1 ± 10 |
| Orgueil (JAXA) | CI1 | 18 | 2 ± 4 | 22 ± 7 | 60 ± 13 | 70 ± 23 | 25 ± 11 | -6 ± 16 |
| Alais (JAXA) | CI1 | 20 | 2 ± 3 | 27 ± 7 | 82 ± 10 | 85 ± 20 | 30 ± 10 | 1 ± 13 |
| Ivuna | CI1 | 22 | -1 ± 2 | 22 ± 3 | 62 ± 7 | 68 ± 11 | 22 ± 5 | -3 ± 7 |
| Murchison (JAXA) | CM2 | 16 | -14 ± 4 | 11 ± 8 | 43 ± 15 | 35 ± 24 | -2 ± 12 | 10 ± 17 |
| MET 01070 | CM1 | 16 | -10 ± 2 | 10 ± 6 | 26 ± 11 | 32 ± 18 | 0 ± 7 | -4 ± 11 |
| NWA 6015 | CO3 | 16 | -7 ± 3 | 11 ± 7 | 22 ± 12 | 33 ± 21 | 4 ± 10 | -10 ± 11 |
| NWA 5933 | CO3 | 16 | -9 ± 3 | 4 ± 8 | 18 ± 16 | 12 ± 24 | -6 ± 11 | 7 ± 14 |
| DaG 136 | CO3 | 16 | -6 ± 4 | 13 ± 8 | 28 ± 12 | 41 ± 23 | 8 ± 11 | -11 ± 14 |
| Allende (JAXA) | CV3 | 16 | -10 ± 4 | 15 ± 10 | 36 ± 17 | 46 ± 30 | 5 ± 14 | -8 ± 15 |
| GRA 06100 | CR1 | 16 | -17 ± 5 | 8 ± 10 | 27 ± 13 | 26 ± 31 | -8 ± 15 | 2 ± 23 |
| Acfer 139 | CR2 | 18 | -16 ± 3 | 14 ± 7 | 35 ± 13 | 43 ± 21 | -2 ± 10 | -6 ± 12 |
| Acfer 182 | CH3 | 12 | -16 ± 5 | 12 ± 11 | 28 ± 22 | 38 ± 35 | -4 ± 16 | -9 ± 17 |
| Tagish Lake (JAXA) | C2-ung | 16 | -11 ± 4 | 16 ± 8 | 44 ± 14 | 50 ± 25 | 5 ± 12 | -3 ± 13 |
| Tagish Lake | C2-ung | 17 | -11 ± 3 | 13 ± 7 | 33 ± 11 | 42 ± 22 | 3 ± 10 | -7 ± 13 |
| Tagish Lake mean | | | -11 ± 3 | 15 ± 5 | 37 ± 8 | 45 ± 16 | 4 ± 8 | -5 ± 9 |
| Tarda (JAXA) | C2-ung | 16 | -14 ± 3 | 19 ± 5 | 52 ± 15 | 60 ± 17 | 5 ± 8 | -6 ± 12 |
| Tarda | C2-ung | 19 | -13 ± 3 | 10 ± 8 | 24 ± 13 | 31 ± 25 | -3 ± 11 | -6 ± 16 |
| Tarda mean | | | -14 ± 2 | 16 ± 4 | 36 ± 10 | 51 ± 13 | 3 ± 6 | -6 ± 9 |

## Discussion

### *Origin of Ni isotope anomalies*

For some elements, mass-independent isotope variations can be due to cosmic ray exposure (CRE), incomplete correction for mass-dependent isotope fractionation, non-mass-dependent isotope fractionation, or radioactive decay. However, for Ni, none of these four processes can account for the observed isotope differences between CI chondrites/Ryugu and the other carbonaceous chondrites. First, CRE effects on Ni isotopes are minor to absent even for samples with CRE ages of several hundreds of millions of years (Ma) (*31*, *32*). Thus, they cannot account for the observed Ni isotope variations among the carbonaceous chondrites given that all of them have much lower

CRE ages. Second, although mass-dependent Ni isotope compositions have not been measured as part of this study, prior studies have shown that the Ni isotope compositions of carbonaceous chondrites vary by only ~0.15 ‰ per atomic mass unit (*30*), which would result in apparent mass-independent effects on $\mu^{60}Ni_{61/58}$, $\mu^{62}Ni_{61/58}$, and $\mu^{64}Ni_{61/58}$ of less than 1, –3, and –11 parts-per-million (ppm), respectively (*33*). These variations are smaller than the observed Ni isotope differences between CI chondrites/Ryugu and the other carbonaceous chondrites, and also than the differences between NC and CC meteorites. Third, prior studies have shown that the correlated nature of Ni isotope variations among meteorites is inconsistent with non-mass-dependent fractionation, such as would for instance be expected for the nuclear field shift effect (*34*). Fourth, the elevated $\mu^{60}Ni$ of CI chondrites/Ryugu cannot result from radioactive decay of short-lived $^{60}Fe$ because, given the relatively low solar system initial $^{60}Fe/^{56}Fe$ (*33*), there is insufficient Fe/Ni fractionation among the carbonaceous chondrites (*35*). Thus, the distinct Ni isotope compositions of CI chondrites and Ryugu, compared to other chondrites, are nucleosynthetic in origin and, as such, confirm the genetic link between CI chondrites and Ryugu that was previously identified based primarily on $^{50}Ti$, $^{54}Cr$, and $^{54}Fe$ isotope anomalies (*9*, *10*, *18*).

The samples in this study, together with meteorites analyzed in prior studies, define a single correlation of $\mu^{64}Ni_{61/58}$ versus $\mu^{62}Ni_{61/58}$ with a slope of ~3 (Fig. 1B). This is the expected slope for a Ni isotope anomaly that is predominantly on $^{58}Ni$ (*30*, *31*). When the Ni isotope data are normalized to a fixed $^{62}Ni/^{61}Ni$, the CC meteorites display higher $\mu^{58}Ni_{62/61}$ values than NC meteorites (Fig. 1C), consistent with the idea that the anomalies are predominantly on $^{58}Ni$. For the $^{62}Ni/^{61}Ni$ normalization, NC and CC meteorites also show correlated $\mu^{58}Ni_{62/61}$ and $\mu^{60}Ni_{62/61}$ variations, where in particular the NC meteorites exhibit a well-defined $\mu^{58}Ni_{62/61}$ versus $\mu^{60}Ni_{62/61}$ slope of 2.03±0.20 (Fig. 1C). Although correlated $\mu^{58}Ni_{62/61}$ and $\mu^{60}Ni_{62/61}$ variations with a slope of ~2 would be consistent with anomalies solely in the normalizing isotope $^{61}Ni$, this would result in a $\mu^{64}Ni_{61/58}$ versus $\mu^{62}Ni_{61/58}$ slope of ~1.5 and not ~3 as observed. Moreover, anomalies in $^{61}Ni$ would also lead to variations in $\mu^{64}Ni_{62/61}$, but this is not observed (Table 1). Thus, the new Ni isotope data, together with data from prior studies, are consistent with the nucleosynthetic Ni isotope heterogeneity predominantly being caused by variations of $^{58}Ni$ and $^{60}Ni$. The correlated $\mu^{58}Ni_{62/61}$ and $\mu^{60}Ni_{62/61}$ variations within each reservoir may reflect variations of Ni produced in the slow neutron capture process (*s*-process) in asymptotic giant branch (AGB) stars (*36*), consistent with the finding of a $^{58}Ni$- and $^{60}Ni$-depleted isotopic composition in an acid-resistant residue of the CI chondrite Ivuna, which is known to be enriched in *s*-process material (*29*). By contrast, the larger $\mu^{58}Ni_{62/61}$ and $\mu^{60}Ni_{62/61}$ values of CC meteorites compared to NC meteorites cannot be explained in this manner, but more likely reflect a higher relative abundance of Ni isotopes produced by nuclear statistical equilibrium either in type Ia supernovae or in the Si/S shell of core-collapse supernovae, which produce an overabundance of $^{58}Ni$ (*37*).

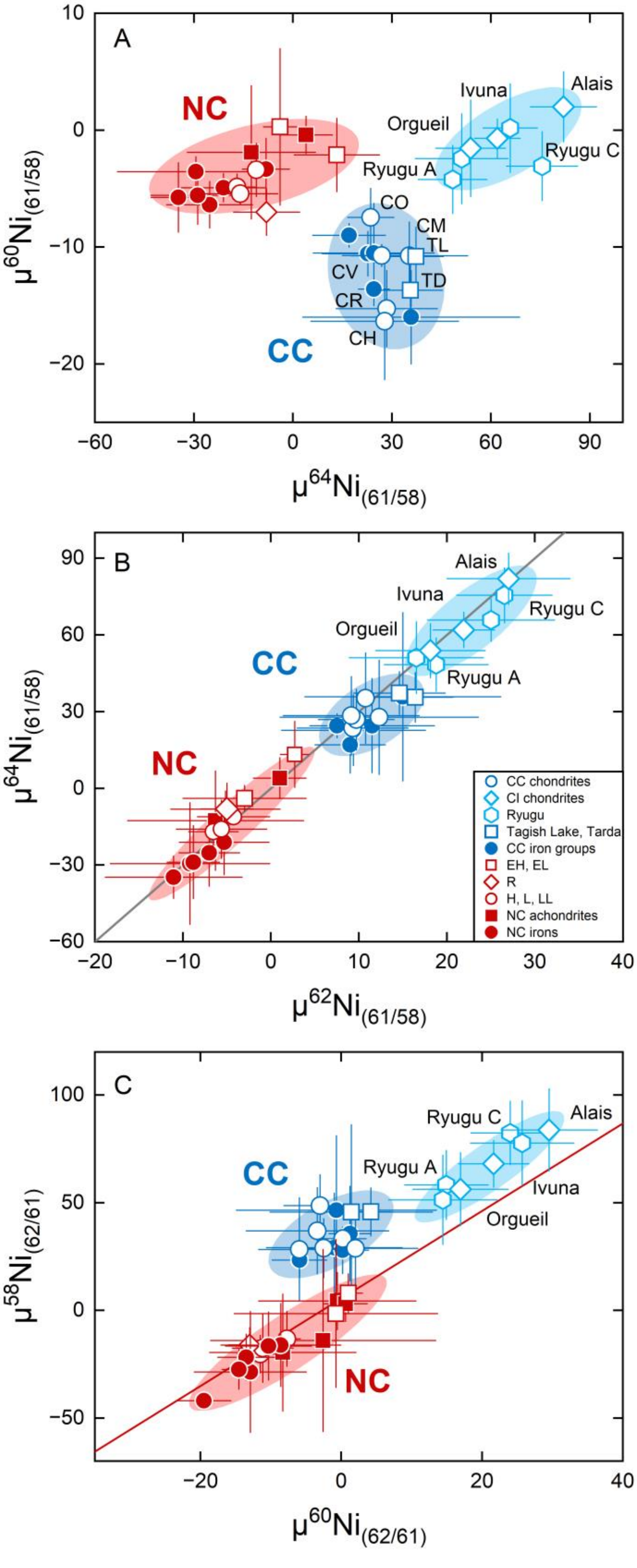

**Figure 1. Nickel isotope anomalies in Ryugu and CI chondrites compared to other meteorites.** (**A**) $\mu^{60}$Ni versus $\mu^{64}$Ni, (**B**) $\mu^{64}$Ni versus $\mu^{62}$Ni, (**C**) $\mu^{58}$Ni versus $\mu^{60}$Ni. Subscripts on axis labels indicate internal normalization to either $^{61}$Ni/$^{58}$Ni (**A**, **B**) or $^{62}$Ni/$^{61}$Ni (**C**). Ryugu and CI chondrites define a distinct compositional cluster in Ni isotope space that is offset from all other carbonaceous chondrites. The grey solid line in (**B**) is a regression through all meteorite data with a slope of ~3 (*31*, *34*). The red solid line in panel (C) is a regression line through all NC meteorites with a slope

of ~2. Data are from this study and compiled from the literature (Table S4). = high-metal enstatite chondrites (EC), EL = low-metal EC, RC = Rumuruti-type chondrites, H = high total iron ordinary chondrites (OC), L = low total iron OC, LL = low total iron - low metallic iron OC.

There are small Ni isotope variations among the CI chondrites and the Ryugu samples, where the two Ryugu A samples (from the first touchdown site) plot close to Orgueil, whereas the two Ryugu C samples (from the second touchdown site) are more similar to Alais and Ivuna (Fig. 1). These variations indicate heterogeneities not only at the scale of individual samples (*18*), but also between the two touchdown sites and individual CI chondrites that each can be linked to one of the two touchdown sites. Isotopic heterogeneities among and between Ryugu samples and CI chondrites have previously been observed for Cr, Ti, and Mo, and were at least in part attributed to the redistribution of isotopically heterogeneous materials during aqueous alteration in the parent body (*18*, *38*). The step-wise dissolution of the CI chondrite Ivuna by acids of increasing strength has revealed that distinct isotopically heterogeneous carriers of Ni exist in CI chondrites (*29*), and so, like for Cr, Ti, and Mo, the redistribution of Ni from these carriers during aqueous alteration in the parent body is a viable mechanism to account for the observed Ni isotope variability among individual Ryugu and CI chondrite samples. Although these heterogeneities are of similar magnitude as the isotopic differences between Ryugu/CI chondrites and other carbonaceous chondrites, the differences are statistically resolved for all Ni isotope ratios (see Supplementary Materials). This is particularly evident in a plot of $\mu^{60}Ni_{61/58}$ versus $\mu^{64}Ni_{61/58}$ (Fig. 1A), where these samples define a distinct cluster that neither overlaps with other CC meteorites nor with NC meteorites.

### *Origin of the unique Fe and Ni isotopic compositions of CI chondrites/Ryugu*

The distinct isotopic compositions of meteorite groups are thought to reflect the accretion of their parent bodies from isotopically heterogeneous dust components, where the distribution of these components in the protoplanetary disk may have varied in space and time. Isotopic variations among the carbonaceous chondrites in particular have been attributed to variable proportions of three main components having distinct isotopic compositions: refractory inclusions [i.e., Ca-Al-rich inclusions (CAIs) and amoeboid olivine aggregates (AOAs)], chondrules, and CI chondrite-like matrix (*24*–*26*). Such a mixing model readily accounts for systematic variations of (*i*) volatile element contents (*24*), (*ii*) mass-dependent isotope compositions of moderately volatile elements (*26*, *27*), (*iii*) O isotope anomalies (*24*, *39*), and (iv) nucleosynthetic $^{54}Cr$ and $^{50}Ti$ anomalies (*24*, *25*) with the mass fraction of matrix in each carbonaceous chondrite (Fig. 2). In this model, difference in $\mu^{54}Cr$ reflect variations in the abundance of chondrules (or their precursors), while differences in $\mu^{50}Ti$ are due to variations in the abundance of refractory inclusions, relative to matrix (for definition of $\mu^{54}Cr$ and $\mu^{50}Ti$ see Fig. 2). The correlation of $\mu^{54}Cr$ versus $\mu^{50}Ti$ among the carbonaceous chondrites can then be attributed to coupled abundance variations of refractory inclusions and chondrules relative to CI chondrite-like matrix (*25*). To assess whether these processes can also account for the observed $\mu^{60}Ni$ variations, we calculated the expected $\mu^{60}Ni_{61/58}$, $\mu^{50}Ti$, and $\mu^{54}Cr$ variations produced by mixing among refractory inclusions, chondrules, and matrix (Materials and Methods). Consistent with the results of prior studies, we find that the $\mu^{54}Cr$ and $\mu^{50}Ti$ variations can readily be accounted for by mixing chondrules and CAIs with CI chondrite-like matrix (Fig. 3A). These mixtures, however, cannot reproduce the observed Ni isotope variations among the carbonaceous chondrites (Fig. 3B, C). Owing to their low Ni contents, the addition of a few percent of CAIs, sufficient to produce the observed $\mu^{50}$Ti variations, does not change the Ni isotopic composition. Similarly, the addition of AOAs, which may be responsible for some of the observed $\mu^{54}$Cr and $\mu^{50}$Ti variations (*40–42*), has little effect on the Ni isotopic compositions and would result in more positive $\mu^{60}Ni_{61/58}$ values in non-CI chondrites and not the observed lower values (Fig. 3B). Mixing between chondrules and CI chondrite-like matrix leads to small $\mu^{60}Ni_{61/58}$ variations, but would also result in higher $\mu^{60}Ni_{61/58}$ values for more matrix-rich carbonaceous chondrites, contrary to the observed more negative $\mu^{60}Ni_{61/58}$ values. As such, mixing between chondrules and matrix could only reproduce the observed Ni isotope variations if the matrix is characterized by a $\mu^{60}Ni_{61/58}$ value of around –20 (for a CI chondrite-like $\mu^{54}$Cr and $\mu^{50}$Ti of the matrix), and not ~0 as measured for CI chondrites.

Given that CI chondrites/Ryugu have distinct Fe and Ni isotope compositions, it is conceivable that these isotope variations predominantly reflect the heterogeneous distribution of FeNi metal. Carbonaceous chondrites contain different types of FeNi metal, which formed by different processes and presumably from different precursor materials. First, FeNi metal grains can be found in AOAs and most likely formed by condensation together with their host AOAs (*43*). Second, the most common FeNi metal in carbonaceous chondrites occurs in association with chondrules, suggesting that this metal was originally part of and, hence, is genetically related to chondrules (*24*). Third, small FeNi metal grains occur in the matrix of some primitive carbonaceous chondrites, but these grains seem to predominantly represent chondrule metal, although some grains may have a condensation origin similar to FeNi metal from AOAs (*44*). Last, FeNi metal in Bencubbin-type (CB) and high-metal (CH) chondrites is thought to have formed by condensation from an impact-generated vapor plume, relatively late in solar system history, at ~4–5 Ma after CAI formation (*45–47*). This metal, therefore, formed later than most of the other carbonaceous chondrites had formed and cannot represent the FeNi metal that was present throughout the accretion disk. As such, CB and CH metal cannot represent the carrier of the Ni isotope variations among the carbonaceous chondrites.

It has been proposed that metal from chondrules is responsible for the $^{54}$Fe isotopic variations observed among carbonaceous chondrites (*23*). Given the correlated nature of the $\mu^{60}Ni_{61/58}$ and $\mu^{54}$Fe variations (Fig. 3D), this metal would then also be responsible for the Ni isotope variations among the carbonaceous chondrites. However, chondrule metal would have the same isotopic composition as the chondrules themselves, but as shown above, mixing between a chondrule component and CI chondrite-like matrix cannot account for the correlated $\mu^{60}Ni_{61/58}$ and $\mu^{54}$Cr variations (Fig. 3C). Instead, the correlation of $\mu^{60}Ni_{61/58}$ with $\mu^{50}$Ti (except for CI chondrites and Ryugu) suggests that the abundance of the FeNi metal is correlated with the abundances of refractory inclusions, the main carrier of the $^{50}$Ti anomalies (Fig. 3B). Since refractory inclusions are characterized by $^{50}$Ti excesses ($\mu^{50}$Ti ~900), the positive correlation of $\mu^{60}Ni_{61/58}$ versus $\mu^{50}$Ti suggests that the FeNi metal is also characterized by $^{60}$Ni excesses. The only known chondrite component having $^{60}$Ni excesses are CAIs ($\mu^{60}Ni_{61/58}$ ~60; Supplementary Materials), but as noted above, they contain too little Ni to significantly modify the Ni isotope composition of bulk carbonaceous chondrites.

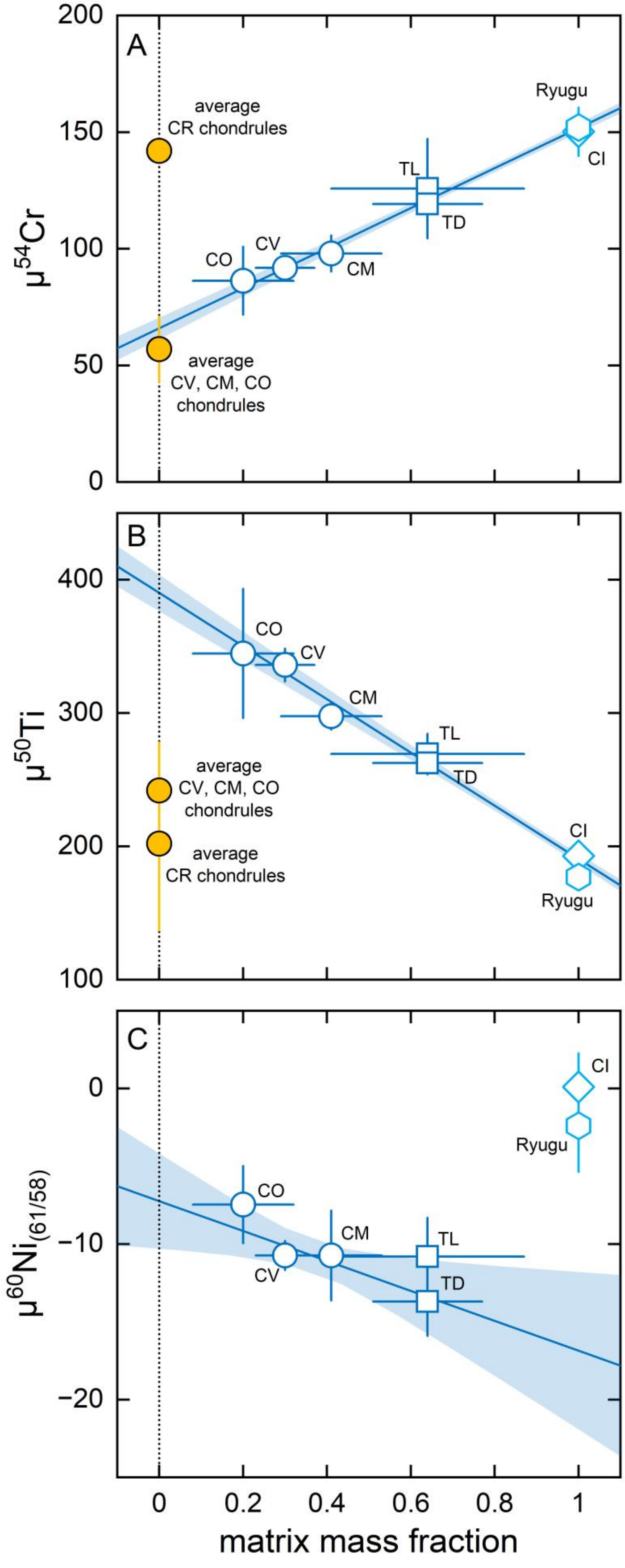

**Figure 2. $^{54}$Cr, $^{50}$Ti, and $^{60}$Ni anomalies versus the matrix mass fractions in carbonaceous chondrites.** (**A**) For $\mu^{54}$Cr all carbonaceous chondrites except CR chondrites define a linear trend with CI chondrites/Ryugu defining one end, and the measured average composition of chondrules from CV, CM, and CO chondrites defining the other end of this trend. Accordingly, the variable $\mu^{54}$Cr values of carbonaceous chondrites are interpreted to result from variable mixing of chondrules and CI chondrite-like matrix. (**B**) For $\mu^{50}$Ti, the same systematic can be observed, but in this case, the y-intercept is not equal to the chondrule composition, because the anomaly in $\mu^{50}$Ti is dominated by refractory inclusions (i.e., CAIs and AOAs). Nevertheless, their abundance is coupled to that of chondrules, indicating a higher fraction of chondrules and CAIs/AOAs in matrix-poor samples. (**C**) Although the absolute variability in $\mu^{60}$Ni versus matrix mass fraction is much smaller, a reasonable correlation ($r^2$ = 0.64) is still observed. However, CI chondrites/Ryugu plot above the trend defined by the other carbonaceous chondrites, indicating that the Ni isotopic composition of CI chondrites/Ryugu is distinct from the expected composition of the matrix in other carbonaceous chondrites. Matrix mass fractions are from (*26*). The mean $\mu^{54}$Cr and $\mu^{50}$Ti values of CV, CM, and CO chondrules are averages of all chondrule measurements (*25*, *42*, *73*, *75*–*79*), which are indistinguishable for all three chondrite groups and as such represent a good average composition for chondrules in these carbonaceous chondrites. CH and CB chondrites are not shown, because they likely formed by impact-related processes and so the matrix in these samples does not represent the original matrix from the solar nebula (*45*–*47*). The Cr and Ti data are reported in the *μ*-notation, i.e., the ppm-deviations from the terrestrial $^{54}$Cr/$^{52}$Cr and $^{50}$Ti/$^{47}$Ti ratios internally normalized to $^{50}$Cr/$^{52}$Cr and $^{49}$Ti/$^{47}$Ti, respectively.

Amoeboid olivine aggregates condensed at slightly lower temperatures than CAIs and have the same Ti and Cr isotopic signatures as CAIs, indicating that less refractory materials having a CAI-like isotopic composition exist (*40*, *41*). The inferred condensation temperature of AOAs of ~1230–1380 K (*43*) overlaps with the condensation temperature of FeNi metal of ~1350 K (*48*), such that FeNi metal would be expected to condense alongside AOAs. Most AOAs from reduced (i.e., less altered) CV chondrites contain some FeNi metal grains (*43*), but chemically AOAs are depleted in Fe and Ni relative to lithophile elements having similar condensation temperatures (i.e., Mg, Si). This has been attributed to the loss of a FeNi metal component, remnants of which are still present in AOAs, perhaps related to aerodynamic sorting of silicate and metal grains during the assembly of AOAs (*43*). Combined, the common isotopic composition of CAIs and AOAs, the similar condensation temperatures of AOAs and FeNi metal, the presence of residual FeNi metal grains inside AOAs, and the chemical evidence for the loss of FeNi metal from AOAs suggest strongly that FeNi metal with a CAI/AOA-like isotopic composition existed and had formed contemporaneously with CAIs and AOAs.

To assess how the heterogeneous distribution of FeNi metal with an assumed CAI/AOA-like $\mu^{60}Ni_{61/58}$ and a Ni abundance as predicted by condensation models (*24*) affects the Ni isotopic composition of carbonaceous chondrites, we included this FeNi metal as a fourth component in the mixing model from above. The calculations show that variations in the abundance of FeNi metal among the bulk carbonaceous chondrites of only up to 5% by mass are sufficient to produce the observed $\mu^{60}Ni_{61/58}$ variations (Fig. 3), and would also result in variations of the bulk Fe/Si ratios of carbonaceous chondrites of up to ~25%. These estimates depend on the assumed $\mu^{60}Ni_{61/58}$ value of the metal, and assuming a higher value than above would allow for a smaller variation in metal abundance and Fe/Si ratios. Thus, the FeNi metal abundance variations we invoke to account for the observed $\mu^{60}Ni_{61/58}$ heterogeneity may also be sufficient to account for the observed ~15% variations in bulk Fe/Si ratios among the major groups of carbonaceous chondrites (*35*). This is consistent with a weak correlation between $\mu^{60}Ni_{61/58}$ values and Fe/Si ratios observed among the carbonaceous chondrites (Fig. S1), suggesting that the $\mu^{60}Ni_{61/58}$ and bulk elemental Fe/Si variations may indeed be directly related. However, testing this hypothesis will require Ni isotope measurements on the residual FeNi metal grains inside AOAs.

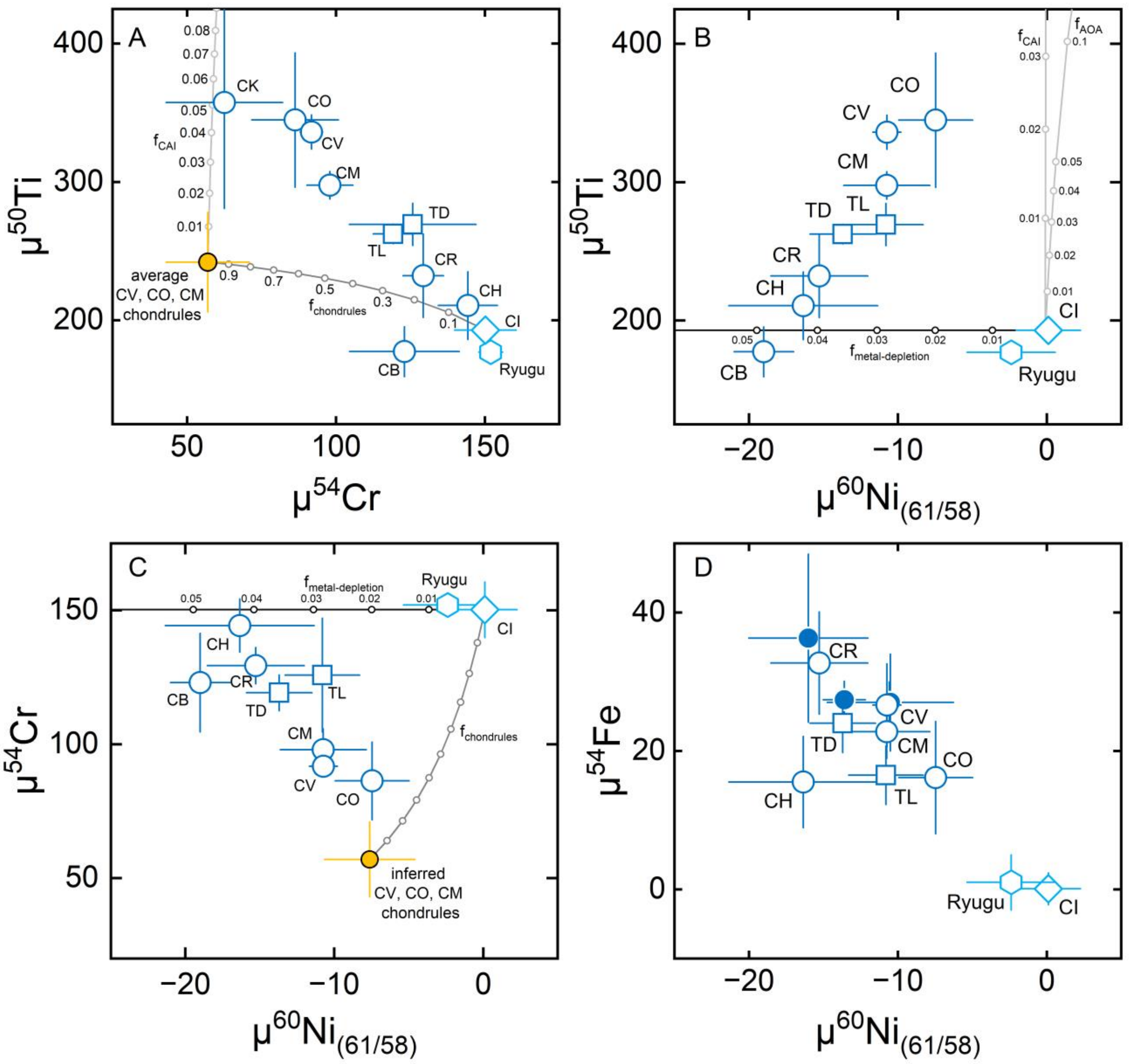

**Figure 3. Relation of Cr, Ti, Ni, and Fe isotope anomalies in carbonaceous chondrites and Ryugu.** Also shown are mixing lines between CI chondrites and CAIs, chondrules, and FeNi metal. Tick marks indicate mass fractions of material added to or lost from (for FeNi metal) CI chondrites. The $\mu^{54}Cr$ and $\mu^{50}Ti$ variations among the carbonaceous chondrites can be accounted for by mixing CI chondrite-like matrix with variable amounts of chondrules and CAIs, respectively (**A**). These mixtures cannot reproduce the $\mu^{60}Ni$ variations, which more likely reflect the heterogeneous distribution (~5 wt%) of FeNi metal characterized by positive $\mu^{60}Ni$ (**B**, **C**). The Ni isotopic composition of the chondrule component in (**C**) is inferred from the non-matrix intercept of Fig. 2C. The linear variations of $\mu^{60}Ni$ and $\mu^{54}Fe$ (**D**) indicate that FeNi metal is likely responsible for both the observed $\mu^{54}Fe$ and $\mu^{60}Ni$ variations. As such, the metal should be characterized by negative $\mu^{54}Fe$ values. Closed symbols in (**D**) represent CC iron meteorite groups. Isotopic data are from the compilation of (*13*) with additional data from (*10*, *18*, *25*, *47*, *80*). $f_{CAI}$, $f_{AOA}$, $f_{metal\text{-}depletion}$, and $f_{chondrules}$ refer to the mass fractions of each component in the mixing calculations.

In our model, the distinct $\mu^{60}Ni_{61/58}$ of CI chondrites/Ryugu compared to all other carbonaceous chondrites reflects the depletion of some FeNi metal. To account for the $\mu^{60}Ni_{61/58}$-$\mu^{50}Ti$ and $\mu^{60}Ni_{61/58}$-$\mu^{54}Cr$ correlations among the non-CI carbonaceous chondrites in this manner requires that this metal depletion systematically varies with the abundances of refractory inclusions and chondrules (which determine the $\mu^{50}Ti$ and $\mu^{54}Cr$ values of the bulk chondrites, respectively) relative to matrix. As noted above, refractory inclusions are strongly depleted in Ni and so cannot be responsible for the observed Ni isotope variations, but chondrules contain sufficient Ni so that variations in their abundance can modify the Ni isotope composition of bulk chondrites. Consequently, the $\mu^{60}Ni_{61/58}$ variations among the non-CI carbonaceous chondrites can be understood as reflecting variable mixtures between chondrules and matrix having CI chondrite-like $\mu^{50}Ti$ and $\mu^{54}Cr$ values but a $\mu^{60}Ni_{61/58}$ value of about –20. The latter is distinct from CI chondrites/Ryugu as a result of FeNi metal depletion (Fig. 3C). The composition of the matrix inferred in this manner overlaps with the $^{60}Ni$, $^{50}Ti$, and $^{54}Cr$ isotopic compositions of the CR, CH, and CB chondrites, yet these chondrites contain relatively low amounts of matrix. However, as noted above, the CH and CB chondrites are thought to have formed in an impact vapor plume (*45*, *46*, *49*), and so their current abundance of matrix almost certainly does not reflect the original matrix fraction in the impactors. Likewise, the low matrix fractions in the CR chondrites have been attributed to recurrent chondrule formation, which resulted in the incorporation of substantial amounts of matrix into the chondrules (*50*). In this case, CR chondrites would originally have contained a high amount of matrix, which is no longer visible petrographically, but is reflected in their isotopic compositions. Together, our results imply that all carbonaceous chondrites, including CI chondrites/Ryugu, formed as mixtures of the same four dust components, namely refractory inclusions, chondrules, FeNi metal, and fine-grained matrix, consistent with the observation that the matrix in carbonaceous chondrites shares many similarities with CI chondrites.

### *Formation of carbonaceous chondrites in a common disk reservoir*

Some current models for the accretion of carbonaceous chondrite parent bodies predict that the variable abundances of refractory inclusions, chondrules, and matrix reflects trapping of these components in a pressure maximum of the disk, which is presumably related to the formation of proto-Jupiter (*25*, *51*) (Fig. 4). In detail, the abundance variations of these components result from the preferential incorporation of refractory inclusions and chondrules over fine-grained matrix (*25*), because planetesimal formation by the streaming instability leads to preferential incorporation of grains with higher Stokes numbers, that is, the largest and densest dust grains. Thus, within the framework of this model, the depletion in FeNi metal grains can be understood if these metal grains were small, such that their incorporation into the carbonaceous chondrite parent bodies was inefficient. This is consistent with the aforementioned occurrence of residual FeNi metal grains, which typically are <10 µm in size, in AOAs (*43*). Such small metal grains would likely be well-coupled to the gas, which makes them difficult to accrete efficiently.

Within this scenario, the uniquely distinct Fe and Ni isotopic compositions of CI chondrites/Ryugu can be attributed to a more efficient accretion of small metal grains compared to all other carbonaceous chondrites. This implies that for CI chondrites/Ryugu the dust enrichment necessary for planetesimal formation occurred by a different process than for other carbonaceous chondrites. Although the gas density of the disk is expected to have varied locally due to, for instance, the interaction with the forming giant planets, this process probably does not result in sufficient dust enrichment (*52*, *53*). The interaction between dust grains with magnetic fields in the disk may have been important, especially for FeNi metal grains, but at present no models exist that would allow quantifying these effects (*54*). Photoevaporation of the gas towards the end of the disk's lifetime has been shown to result in sufficient dust enrichment to trigger planetesimal formation (*55*). Importantly, this process results in the accretion of all the available dust without prior fractionation among the different dust components and, as such, is different from the aforementioned dust pile-up in a pressure maximum that appears to have been important for the accretion of the non-CI carbonaceous chondrites. Thus, the formation of CI chondrites/Ryugu by dust enrichment through photoevaporation of the gas can account for both the more efficient accretion of FeNi metal grains (because all available dust is accreted) and for the distinct isotopic compositions of CI chondrites compared to all other carbonaceous chondrites (because these two groups of chondrites formed by different processes).

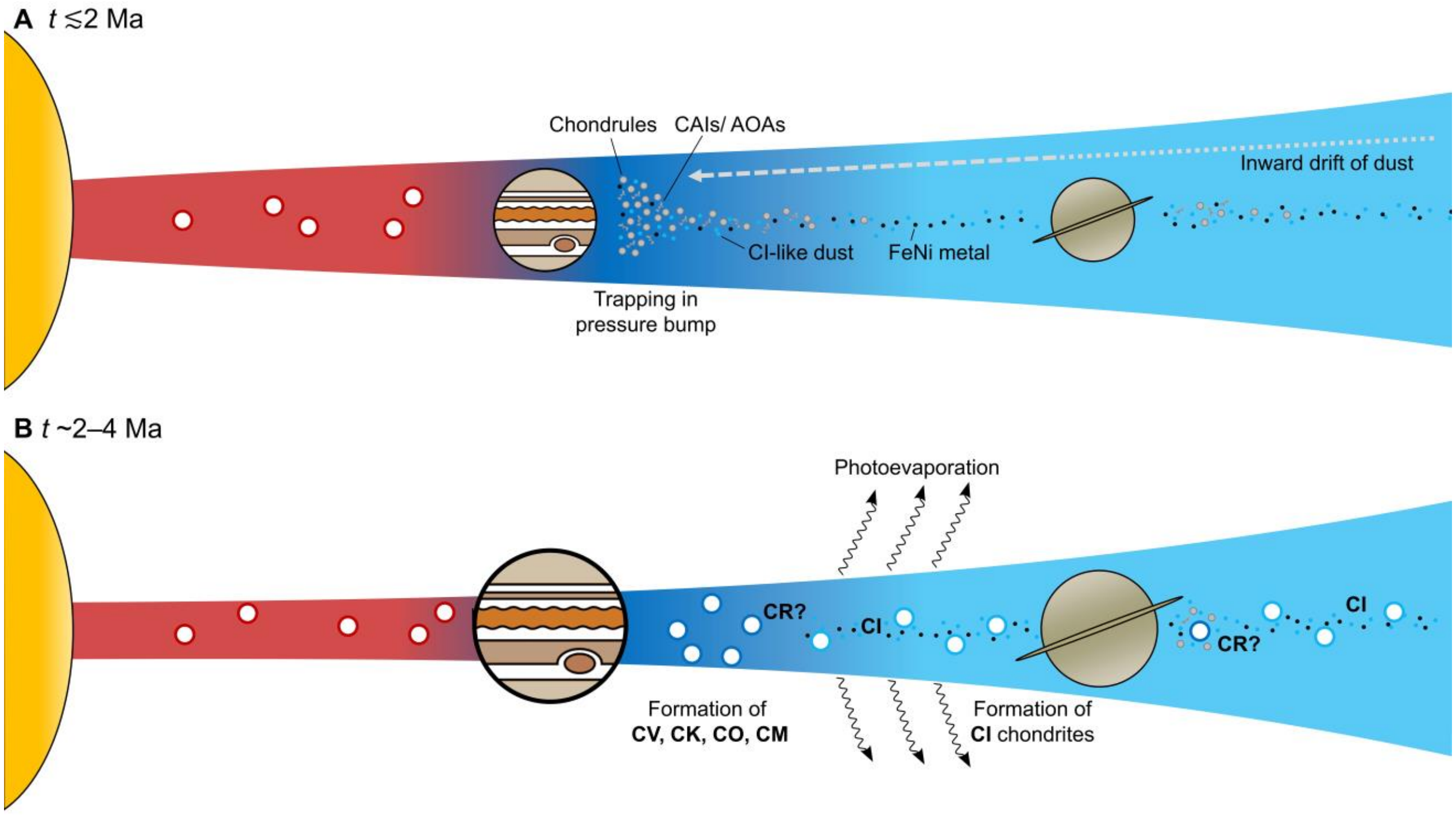

**Figure 4. Cartoon illustrating the formation mechanism and accretion region of carbonaceous chondrite parent bodies.** (**A**) The four major components of carbonaceous chondrites (chondrules, refractory inclusions, matrix, and FeNi metal) are transported through the disk towards the Sun and are trapped in a pressure maximum, which was presumably located outside the orbit of Jupiter. (**B**) The parent bodies of several non-CI carbonaceous chondrites (i.e., CV, CK, CO, and CM) form in this pressure maximum and incorporate different proportions of refractory inclusions, chondrules, FeNi metal, and fine-grained matrix. Owing to their small size, accretion of FeNi metal grains is inefficient in all these bodies. At the end of the disk's lifetime, nebular gas is removed via photoevaporation leading to a late burst of planetesimal formation over a wider area of the disk. CI chondrites and Ryugu form by this process and incorporate the entire background population of dust in the disk, and so accrete the small FeNi metal grains more efficiently than the other carbonaceous chondrites. CR chondrites form at about the same time but by a similar process as the

other non-CI carbonaceous chondrites. The original formation location of the CR chondrites is unclear.

Gas removal by photoevaporation only occurs close to the end of the disk's lifetime. For the solar system, dispersal of the gaseous disk has been estimated to have occurred at ~4 Ma after solar system formation (*56*). This is consistent with an accretion age of around 4 Ma after solar system formation (*57*) inferred based on $^{53}Mn$-$^{53}Cr$ ages for carbonates from CI chondrites and Ryugu of ~4–6 Ma after solar system formation (*9*, *57*, *58*). However, two recent studies reported older $^{53}Mn$-$^{53}Cr$ ages for Ryugu carbonates of ~2.6 Ma (*59*) and <1.8 Ma (*60*) after solar system formation, which have been interpreted to reflect a relatively early accretion of the Ryugu parent body. All these carbonate ages were determined by secondary ion mass spectrometry (SIMS) and the differences among them have been attributed to the use of different standard materials, resulting in uncertainty about the veracity of these ages. However, recent $^{53}Mn$-$^{53}Cr$ studies on different Ryugu rock fragments utilizing inductively coupled plasma mass spectrometry (*18*) or thermal ionization mass spectrometry (*61*), which compared to SIMS are both less sensitive to the standard material used, found ages between ~4 and ~5 Ma after solar system formation. These ages are consistent with most of the previously reported carbonate ages and with a relatively late formation of Ryugu and CI chondrites towards the end of the disk's lifetime.

Some other carbonaceous chondrites such as the CM (*62*) and CR chondrites (*63*, *64*) have similar inferred accretion ages as the CI chondrites. This implies that different carbonaceous chondrite parent bodies formed about contemporaneously but by different processes. Accretion of non-CI carbonaceous chondrites by trapping dust in a pressure maximum (*25*, *51*) would likely occur as long as the gaseous disk persisted. As such, at the end of the disk‘s lifetime, the dust enrichment necessary for planetesimal formation may have occurred via different mechanisms and at different locations in the disk implying that CI chondrites formed outside the pressure maximum in which the non-CI carbonaceous chondrites formed (Fig. 4). Thus, our results allow for the formation of all carbonaceous chondrites, including CI chondrites/Ryugu, from the same dust components in the same region of the disk instead of distinct reservoirs separated by the giant planets (*10*, *21*).

Since photoevaporation of the gas likely triggered planetesimal formation over a wide range of heliocentric distances and is not expected to have resulted in substantial separation of distinct dust grains, our model predicts that objects formed at even greater heliocentric distance than the CI chondrites, such as Kuiper Belt objects and Oort Cloud comets, have the same non-volatile element isotopic composition as CI chondrites. A further prediction of our model is that there should be no differentiated meteorites having CI chondrite-like Fe and Ni isotopic compositions. This is because $^{26}Al$, the dominant heat source for melting and differentiation of planetesimals, was sufficiently abundant only within the first ~1.5 Ma of the solar system (*65*). Consequently, planetesimals that formed by dust enrichment through photoevaporation at the end of the disk's lifetime contained too little $^{26}Al$ to melt. This expectation is consistent with available Fe and Ni isotopic compositions for CC-type iron meteorites, all of which are distinct from CI chondrites (*31*, *33*, *34*, *66*). However, further analyses of a more comprehensive set of samples representing a larger number of parent bodies are necessary to test this hypothesis.

## Materials and Methods

### *Samples and sample preparation*

A total of ~5.4 g of sample from asteroid Ryugu collected during two touchdowns were returned by the Hayabusa2 spacecraft (*17*, *67*). Two subsamples from Chamber A (A0106, A0106-A0107) and two from Chamber C (C0107, C0108) were digested for isotopic analyses (Table 1). In addition, six carbonaceous chondrite powders labeled JAXA (Table 1: Orgueil, Alais, Murchison, Allende, Tagish Lake, and Tarda) were digested and processed alongside the Hayabusa2 samples. Sample digestion for these samples was conducted at Tokyo Tech. Powder aliquots were digested using mixtures of HF-$HNO_3$-HCl-$H_2O_2$ on a hot plate and with ultrasonic agitation. Approximately 80% of the solutions were taken for sequential separation of several elements for isotopic analyses. For the present study, we also measured 11 additional carbonaceous chondrite samples. Of these, some were digested for this study, while others were elution cuts from previous studies. The masses digested, the original masses of the homogenized sample powders, and details on the processing history for each sample are summarized in Table S1. The chemical separation and purification of Ni from all samples followed our previously established protocols, which involve a 3-step ion-exchange chromatographic procedure (*31*). Sample solutions containing approximately 20–200 µg of Ni were loaded in 10 ml of 0.6 M HCl–90% acetone onto BioRad PolyPrep columns filled with 2 ml pre-cleaned and conditioned BioRad AG 50WX4 cation exchange resin (200–400 mesh). Most of the sample matrix (e.g., Fe, Cr) was eluted with an additional 35 ml 0.6 M HCl–90% acetone and 10 ml 0.6 M HCl–95% acetone before Ni was collected in 6 ml 0.6 M HCl–95% acetone–0.1 M dimethylglyoxime (DMG). This was followed by repeated dry downs using concentrated oxidizing acids to remove any organics remaining from the chemical procedure. The main purpose of the second column is to remove Ti (and Zn), which is accomplished using an anion exchange column procedure using BioRad AG1 X8 resin (100–200 mesh). The Ni fractions were loaded in 2 ml 0.5 M HF–1 M HCl and eluted with an additional 7 ml 0.5 M HF–1 M HCl. Finally, a third column containing BioRad AG MP-1 X4 anion exchange resin (100–200 mesh) was used to remove residual Zn and Fe. For this column, the Ni cuts from the second chemistry were re-dissolved in 1 ml 6 M HCl–0.01% $H_2O_2$ and again eluted in 5 ml 6 M HCl–0.01% $H_2O_2$. The final Ni solution was evaporated to dryness, re-dissolved multiple times in concentrated $HNO_3$ and diluted to 0.3 M $HNO_3$ for analysis by MC-ICP-MS. The overall yield of the chemical procedure is ~90 %. Total procedural blanks for the Ni separation and purification were <10 ng Ni and, hence, negligible given the amount of Ni analyzed for each sample. This procedure achieves sufficiently low $^{58}Fe/^{58}Ni$ ($\leq 4.2 \times 10^{-6}$) and $^{64}Zn/^{64}Ni$ ($\leq 1.2 \times 10^{-3}$) ratios in the final purified Ni cuts to allow for accurate and precise correction of isobaric interferences (*31*, *68*).

### *Nickel isotope measurements*

All isotope measurements were performed on the Thermo Scientific Neptune Plus MC-ICP-MS at the Institut für Planetologie, University of Münster, using our established measurement protocol (*31*). Only Ivuna was measured at the Max Planck Institute for Solar System Research in Göttingen using a Thermo Scientific Neoma MC-ICP-MS, using the same measurement protocol as in Münster. Its accuracy is testified by the good agreement between the Ni isotopic composition of Ivuna determined in this study and that reported previously (*29*). The purified Ni cuts were dissolved in 0.3 M $HNO_3$ and introduced into the mass spectrometer using a Savillex C-Flow PFA nebulizer connected to a Cetac Aridus II desolvator at an uptake rate of ∼50 μL/min. Measurements were made on a flat top section of the left-peak shoulder using medium mass resolution mode to avoid possible interferences on mass $^{57}$Fe from $^{40}$Ar$^{16}$OH or $^{40}$Ar$^{17}$O and $^{132}$Xe$^{2+}$ on $^{66}$Zn (*29*, *33*). Sample and standard solutions were measured at a concentration of approximately 1.7 µg/ml using a combination of standard sampler and X-skimmer cones yielding intensities of ~90V on $^{58}$Ni. The Ni concentrations of the sample solutions were matched to within 5 % of the SRM 986 solution standard. Ion beams at masses 58, 60, 61, 62, and 64 were simultaneously collected in a single cycle. In addition to the Ni masses, ion beams at masses 57 and 66 were also monitored to correct for potential isobaric interferences from $^{58}$Fe and $^{64}$Zn. All ion beams were collected in Faraday cups connected to amplifiers with $10^{11}$ Ω feedback resistors, except for $^{58}$Ni ($10^{10}$ Ω resistor) as well as $^{57}$Fe and $^{66}$Zn ($10^{12}$ Ω resistors). Before each sample or standard measurement, baselines were measured as on peak zeros (OPZ) for 20 × 8.4 s using the same acid solution used for the sample and standard solutions. Sample and standard measurements consisted of 50 cycles of 8.4 s integrations each. Instrumental mass bias was corrected by internal normalization to either $^{61}$Ni/$^{58}$Ni ('1/8') = 0.016744 or $^{62}$Ni/$^{61}$Ni ('2/1') = 3.1884 using the exponential law (*69*). All data are reported in $\mu^{i}$Ni values (i.e., ppm-deviation from the mean value of the terrestrial SRM 986 solution standard analyzed bracketing the sample measurements). For samples analyzed several times, reported $\mu$-values represent the mean of pooled solution replicates, and uncertainties are reported as two standard errors (2 s.e.). The accuracy and precision of the isotope measurements were assessed by repeated analyses of the NIST 361 metal. The external reproducibility (two standard deviations, 2 s.d.) of the isotope measurements of NIST 361 are 5, 9, 18, 28, 13, and 18 ppm for $\mu^{60}$Ni (1/8), $\mu^{62}$Ni (1/8), $\mu^{64}$Ni (1/8), $\mu^{58}$Ni (2/1), $\mu^{60}$Ni (2/1), and $\mu^{64}$Ni (2/1), respectively (Table S2).

### *Four-component mixing model*

Variations in the isotopic compositions of carbonaceous chondrites reflect variable proportions of the four major chondrite components, namely CI chondrite-like matrix, chondrules (or chondrule precursors), refractory inclusions (CAIs, AOAs), and FeNi metal. These components have distinct isotopic compositions. Previous studies have shown that the variation in $\mu^{54}$Cr–$\mu^{50}$Ti space can be understood as mixtures of CI chondrite-like matrix with chondrules and CAIs/AOAs, where the abundances of the last two components are coupled to each other (*25*). We build on this observation and calculate mixing lines between CI chondrite-like matrix and the other chondrite components to evaluate the effect of variations in the abundance of FeNi metal on the isotopic compositions of the carbonaceous chondrites (Fig. 3). The chemical and isotopic compositions of the components used in the mixing calculations are summarized in Table S3.

We assume the matrix to be chemically and isotopically like CI chondrites. We note that the chemical composition of the matrix measured today may be distinct from CI chondrites (*70*), which is presumably due to post-accretionary modifications by parent body (aqueous alteration, thermal metamorphism) and terrestrial (weathering) processes. Furthermore, the matrix in carbonaceous chondrites may itself be a mixture of CI chondrite-like material with a 'chondrule-related' matrix, whose chemical composition is fractionated as a result of chondrule formation (*71*). As such, the non-CI chondrite-like chemical composition of the matrix in some carbonaceous chondrites is not inconsistent with the observation that variations in the bulk chemical and isotopic composition of the different carbonaceous chondrite groups can be attributed to varying abundances of CI chondrite-like matrix. For the chondrule component, we used the average chemical composition of CV, CM, CO, and Karoonda-type (CK) chondrules from the ChondriteDB (*72*) combined with their mean Cr and Ti isotopic compositions (e.g. as compiled in (*25*, *42*, *73*)). Although the isotopic compositions of individual chondrules vary, mean compositions of chondrules from different CC chondrite groups (except the CR chondrites) are indistinguishable from each other (e.g. *25*), indicating that these chondrite groups contain the same population of chondrules with a well-defined average isotopic composition (Table S3). The Ni isotopic composition of chondrules has not yet been measured. We, therefore, assume this composition to be that of the inferred non-matrix component (Fig. 2C). Last, as discussed before, the FeNi metal is assumed to almost exclusively consist of these two elements (*24*) and has a CAI-like Ni isotopic composition. This is consistent with condensation models, which predict this metal to almost exclusively consists of the common siderophiles Fe, Ni, and Co with trace amounts of Pd and Rh (due to their lower 50% condensation temperature), while the more refractory siderophiles condensed earlier into refractory metal nuggets (RMN), together with CAIs (*24*, *74*).

## Acknowledgments

Hayabusa2 was developed and built under the leadership of the Japan Aerospace Exploration Agency (JAXA), with contributions from the German Aerospace Center (DLR) and the Centre National d'Études Spatiales (CNES), and in collaboration with NASA, and other universities, institutes, and companies in Japan. The curation system was developed by JAXA in collaboration with companies in Japan. We gratefully acknowledge constructive comments by three anonymous reviewers and helpful discussions with Joanna Drążkowska. **Funding:** This project was funded by the Deutsche Forschungsgemeinschaft (DFG, German Research Foundation) – Project-ID 263649064 – TRR 170. This is TRR 170 pub. no. 213. **Author contributions:** H.Y. and T.Y. coordinated the isotopic analyses of the samples among members of the Hayabusa2-initial-analysis chemistry team. F.S. and T.Y. processed the samples and separated Ni from the matrix. F.S. measured the Ni isotopic composition. F.S. and T.K. wrote the paper, with contributions from all co-authors. **Competing interests:** The authors declare that they have no competing interests. **Data and materials availability:** The authors declare that data supporting the findings of this study are available within the paper and its Supplementary Materials. Data from the Hayabusa2 samples and other data from the mission are available at the DARTS archive at www.darts.isas.jaxa.jp/curation/hayabusa2 and www.darts.isas.jaxa.jp/planet/project/hayabusa2/.

## Supplementary Materials

### *Ni isotopic composition of CAIs*

The $\mu^{54}Fe$ and $\mu^{60}Ni$ values of most CAIs investigated so far are similar to those of their host chondrites, indicating that the original Fe and Ni isotope signatures of the CAIs were modified by alteration on the carbonaceous chondrite parent bodies (*68*, *81*), most likely by oxidation and dissolution of the metal (*82*). Having this issue in mind, previous studies argued that the high-Fe pyroxene mineral separate from the Egg 2 CAI from Allende represents the most pristine Fe and Ni composition for CAIs because of its low FeO content, magnetic susceptibility, and CAI-like Ti and Sr isotopic composition (*68*, *81*). The same conclusion has also been drawn based on Ni isotope measurements on CAIs separated from the reduced CV chondrite Efremovka (*29*), which was subject to less severe alteration than Allende (*83*). All these CAIs display large mass-dependent isotope fractionations, which can potentially result in residual spurious mass-independent isotope effects when using the exponential law for mass bias correction (*33*). To account for this effect, the measured isotope anomalies were corrected using the approach of Tang and Dauphas (2012). Table S5 summarizes the Fe and Ni isotope data of the CAIs which show little effects of parent body alteration. These CAIs have been used to determine the current best estimate of the Fe and Ni isotope composition of CAIs used in this study.

### *Ni isotopic variations among Ryugu, CI, and other CC chondrites*

Given that the number of analyzed samples is small, a Welch's t-test was conducted to evaluate whether there are statistically significant Ni isotopic differences between CI chondrites/ Ryugu (n = 5; Ryugu A, Ryugu C, Orgueil, Alais, Ivuna) and other carbonaceous chondrites (n = 8; CM, CO, CV, CR, CH, CB, TL, TD). The Welch's t-test is generally applied when there is a difference between the variations of two populations and also when their sample sizes are unequal. The designated null hypothesis (H0) is that the average $\mu^{i}Ni$ values of both populations are equal, where *i* stand for 60, 62, or 64 when normalized to $^{61}Ni/^{58}Ni$. If the two-tailed P value is less than the defined significance threshold (α) of 0.05, then H0 can be rejected at 95% confidence. We find that the Welch's t-test yields two-tailed P values $<0.05$ for all three $\mu^{i}Ni$ values, indicating that we can confidently reject the null hypothesis (H0) and conclude that CI chondrites/ Ryugu are isotopically distinct from the other carbonaceous chondrites.

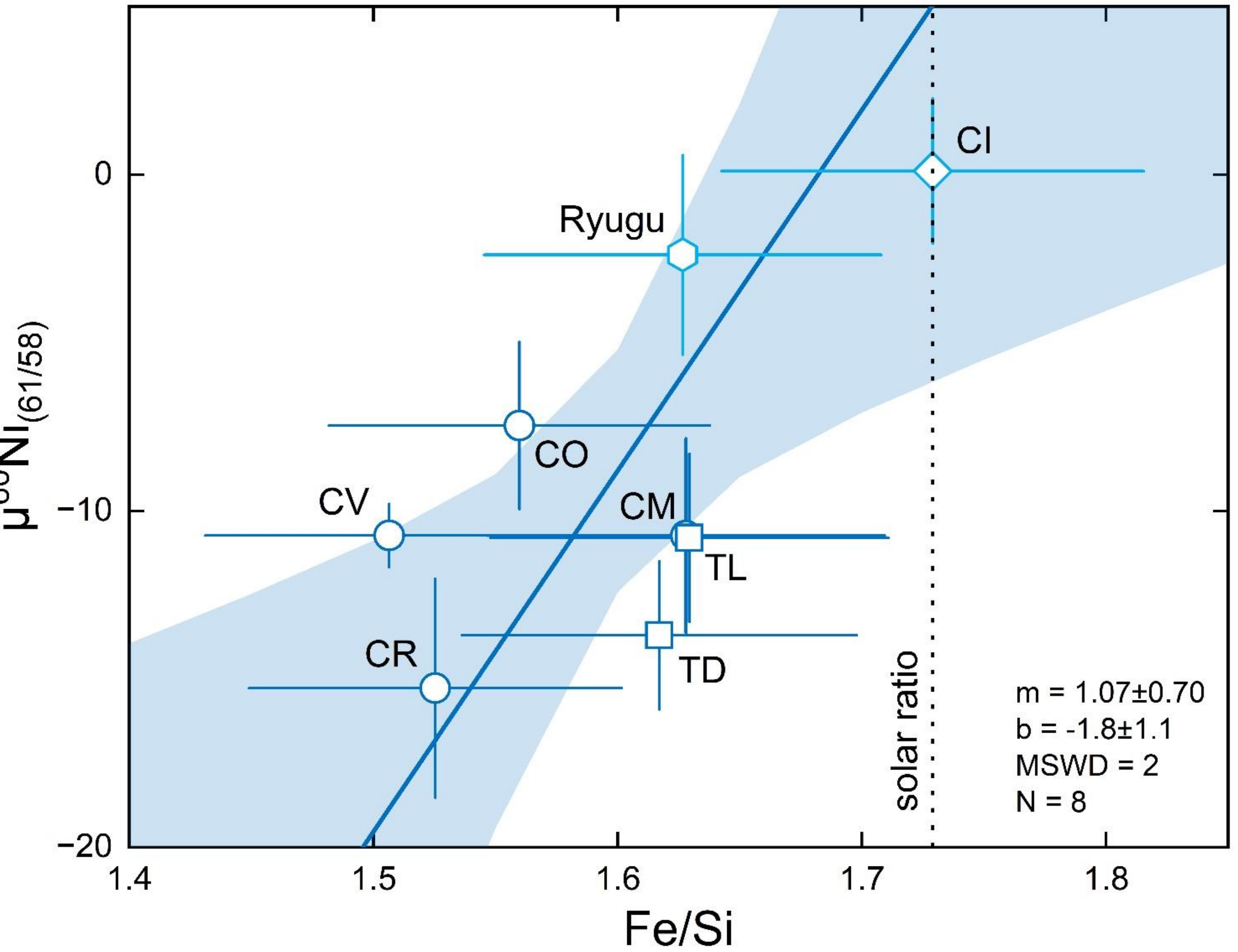

**Figure S1. Diagram of $\mu^{60}Ni_{(61/58)}$ versus Fe/Si mass ratios of carbonaceous chondrites.** The weak correlation suggests that the observed ~15% variations in bulk Fe/Si ratios among carbonaceous chondrites may indeed be directly related to the variation in $\mu^{60}Ni_{61/58}$, which we attribute to abundance variations of ~5 wt% of FeNi metal grains among their parent bodies.

**Table S1. Ni isotopic data of samples investigated in this study.** Uncertainties of individual samples represent two standard errors (2 s.e.), where N is the number of measurements. Data is internally normalized to either $^{61}Ni/^{58}Ni = 0.016744$ or $^{62}Ni/^{61}Ni = 3.1884$. The chemical procedure used at the Tokyo Institute of Technology is described in the methods and (***84***). Matrix aliquots containing the bulk Ni of samples processed at the Institut für Planetologie Münster (IfP) were digested HF-HNO3-HCl-HClO4 mixtures and then separated from Sr or W using cation (AG50W-X8) or anion exchange (AG1-X8) exchange resin, respectively. All samples, regardless of their processing at the Tokyo Institute of Technology or Institut für Planetologie Münster were further processed at IfP to purify Ni.

| Sample | Class | Mass digested (g) | Mass homogenized (g) | Processed? | N | Norm. $^{61}Ni/^{58}Ni$ | | | Norm. $^{62}Ni/^{61}Ni$ | | |
|---|---|---|---|---|---|---|---|---|---|---|---|
| | | | | | | $\mu^{60}Ni$ | $\mu^{62}Ni$ | $\mu^{64}Ni$ | $\mu^{58}Ni$ | $\mu^{60}Ni$ | $\mu^{64}Ni$ |
| Ryugu samples | | | | | | | | | | | |
| A0106–A0107 | | 0.0239 | 0.0289 | Fe, U (Tokyo Tech) | 24 | -4 ± 3 | 19 ± 6 | 48 ± 10 | 58 ± 18 | 15 ± 9 | -7 ± 12 |
| A0106 | | 0.0146 | 0.0175 | Fe, U (Tokyo Tech) | 20 | -2 ± 4 | 17 ± 8 | 51 ± 14 | 51 ± 24 | 14 ± 11 | 2 ± 15 |
| C0108 | | 0.0222 | 0.0333 | Fe, U (Tokyo Tech) | 24 | -3 ± 3 | 27 ± 5 | 76 ± 11 | 82 ± 17 | 24 ± 8 | -3 ± 12 |
| C0107 | | 0.0128 | 0.0174 | Fe, U (Tokyo Tech) | 20 | 0 ± 4 | 25 ± 7 | 66 ± 8 | 78 ± 22 | 26 ± 11 | -8 ± 16 |
| Carbonaceous chondrites | | | | | | | | | | | |
| Orgueil | CI1 | 0.401 | 1.12 | Sr matrix cut (IfP) | 37 | -3 ± 3 | 15 ± 4 | 43 ± 7 | 46 ± 14 | 12 ± 7 | -1 ± 10 |
| Orgueil (JAXA) | CI1 | 0.02 | 0.05 | Fe, U (Tokyo Tech) | 18 | 2 ± 4 | 22 ± 7 | 60 ± 13 | 70 ± 23 | 25 ± 11 | -6 ± 16 |
| Alais (JAXA) | CI1 | 0.022 | 0.051 | Fe, U (Tokyo Tech) | 20 | 2 ± 3 | 27 ± 7 | 82 ± 10 | 85 ± 20 | 30 ± 10 | 1 ± 13 |
| Ivuna | CI1 | 0.05 | 0.1118 | Te chemistry cut (MPS) | 22 | -1 ± 2 | 22 ± 3 | 62 ± 7 | 68 ± 11 | 22 ± 5 | -3 ± 7 |
| Murchison (JAXA) | CM2 | 0.025 | 1.65 | Fe, U (Tokyo Tech) | 16 | -14 ± 4 | 11 ± 8 | 43 ± 15 | 35 ± 24 | -2 ± 12 | 10 ± 17 |
| MET 01070 | CM1 | 0.00457 | 1.7667 | Ni (IfP) | 16 | -10 ± 2 | 10 ± 6 | 26 ± 11 | 32 ± 18 | 0 ± 7 | -4 ± 11 |
| NWA 6015 | CO3 | 0.00487 | >2 | Ni (IfP) | 16 | -7 ± 3 | 11 ± 7 | 22 ± 12 | 33 ± 21 | 4 ± 10 | -10 ± 11 |
| NWA 5933 | CO3 | 0.00486 | 3.239 | Ni (IfP) | 16 | -9 ± 3 | 4 ± 8 | 18 ± 16 | 12 ± 24 | -6 ± 11 | 7 ± 14 |
| DaG 136 | CO3 | 0.00648 | 1.83 | Ni (IfP) | 16 | -6 ± 4 | 13 ± 8 | 28 ± 12 | 41 ± 23 | 8 ± 11 | -11 ± 14 |
| Allende (JAXA) | CV3 | 0.025 | ~4000 (USNM) | Fe, U (Tokyo Tech) | 16 | -10 ± 4 | 15 ± 10 | 36 ± 17 | 46 ± 30 | 5 ± 14 | -8 ± 15 |
| GRA 06100 | CR1 | 0.281 | 0.281 | Sr matrix cut (IfP) | 16 | -17 ± 5 | 8 ± 10 | 27 ± 13 | 26 ± 31 | -8 ± 15 | 2 ± 23 |
| Acfer 139 | CR2 | 0.00538 | 2.09 | Ni (IfP) | 18 | -16 ± 3 | 14 ± 7 | 35 ± 13 | 43 ± 21 | -2 ± 10 | -6 ± 12 |
| Acfer 182 | CH3 | 0.553 | 0.553 | W matrix cut (IfP) | 12 | -16 ± 5 | 12 ± 11 | 28 ± 22 | 38 ± 35 | -4 ± 16 | -9 ± 17 |
| Tagish Lake (JAXA) | C2-ung | 0.025 | 1.06 | Fe, U (Tokyo Tech) | 16 | -11 ± 4 | 16 ± 8 | 44 ± 14 | 50 ± 25 | 5 ± 12 | -3 ± 13 |
| Tagish Lake | C2-ung | 0.486 | 1.5 | Sr matrix cut (IfP) | 17 | -11 ± 3 | 13 ± 7 | 33 ± 11 | 42 ± 22 | 3 ± 10 | -7 ± 13 |
| Tarda (JAXA) | C2-ung | 0.025 | 0.212 | Fe, U (Tokyo Tech) | 16 | -14 ± 3 | 19 ± 5 | 52 ± 15 | 60 ± 17 | 5 ± 8 | -6 ± 12 |
| Tarda | C2-ung | 0.00619 | ~0.5 | Ni (IfP) | 19 | -13 ± 3 | 10 ± 8 | 24 ± 13 | 31 ± 25 | -3 ± 11 | -6 ± 16 |

**Table S2. Ni isotopic data for NIST SRM 361 compared to the literature.**

| Sample | N | Norm. $^{61}Ni/^{58}Ni$ | | | Norm. $^{62}Ni/^{61}Ni$ | | |
|---|---|---|---|---|---|---|---|
| | | $\mu^{60}Ni$ (±2s.e.) | $\mu^{62}Ni$ (±2s.e.) | $\mu^{64}Ni$ (±2s.e.) | $\mu^{58}Ni$ (±2s.e.) | $\mu^{60}Ni$ (±2s.e.) | $\mu^{64}Ni$ (±2s.e.) |
| NIST 361 #1 | 37 | -1 ± 2 | -1 ± 4 | 1 ± 8 | -4 ± 12 | -2 ± 6 | 4 ± 8 |
| NIST 361 #2 | 21 | -2 ± 3 | 7 ± 5 | 26 ± 8 | 22 ± 15 | 5 ± 7 | 4 ± 11 |
| NIST 361 #3 | 10 | 2 ± 5 | 8 ± 8 | 18 ± 17 | 24 ± 26 | 10 ± 12 | -5 ± 17 |
| NIST 361 #4 | 12 | -1 ± 4 | 6 ± 8 | 18 ± 13 | 20 ± 24 | 5 ± 12 | -1 ± 13 |
| NIST 361 #5 | 8 | 3 ± 9 | 11 ± 16 | 16 ± 19 | 35 ± 48 | 15 ± 25 | -17 ± 29 |
| Average (±2 s.d.) | 5 | 0 ± 5 | 6 ± 9 | 16 ± 18 | 19 ± 28 | 7 ± 13 | -3 ± 18 |
| Average literature* (±2 s.d.) | 5 | -1 ± 1 | 5 ± 11 | 8 ± 15 | 16 ± 35 | 4 ± 12 | -7 ± 32 |
| Bulk silicate Earth† | | -1 ± 1 | 4 ± 1 | 12 ± 2 | 11 ± 3 | 3 ± 1 | 1 ± 3 |

Data sources: *(*30*–*32*, *34*) and †(*34*).

**Table S3. Parameters used to calculate mixing lines.** Chemical data for chondrules from ChondriteDB **(*72*)**, for CAIs from **(*85*)**, for FeNi metal and chondrites from **(*24*)**, for Tarda from **(*86*)**, for Ryugu from **(*9, 59*)**, matrix mass fractions (except Tarda) from **(*26*)**, Ni isotopic data as in Tables S3-4. Ti and Cr data were taken from the compilation in **(*13*)** with additions from **(*18, 25, 26, 41*)**.

| Sample | $\varepsilon^{60}Ni_{(61/58)}$ | 95%CI | Fe (wt%) | Ni (wt%) | Ti (ppm) | Cr (wt%) | Si (wt%) | $\varepsilon^{50}Ti$ | 95%CI | $\varepsilon^{54}Cr$ | 95%CI | Matrix mass fraction | 2σ | Fe/Si | 2σ |
|---|---|---|---|---|---|---|---|---|---|---|---|---|---|---|---|
| Ryugu | -0.02 | 0.03 | 19.86 | 1.20 | 473 | 0.27 | 12.21 | 1.77 | 0.07 | 1.52 | 0.04 | 1.00 | 0.01 | 1.63 | 0.08 |
| CI | 0.00 | 0.02 | 18.50 | 1.06 | 440 | 0.26 | 10.70 | 1.93 | 0.06 | 1.50 | 0.10 | 1.00 | 0.01 | 1.73 | 0.09 |
| CM | -0.11 | 0.03 | 21.00 | 1.21 | 619 | 0.31 | 12.90 | 2.98 | 0.10 | 0.98 | 0.08 | 0.41 | 0.12 | 1.63 | 0.08 |
| CO | -0.07 | 0.02 | 24.80 | 1.41 | 765 | 0.36 | 15.90 | 3.45 | 0.49 | 0.86 | 0.15 | 0.20 | 0.12 | 1.56 | 0.08 |
| CV | -0.11 | 0.01 | 23.50 | 1.34 | 899 | 0.36 | 15.60 | 3.36 | 0.12 | 0.92 | 0.04 | 0.30 | 0.07 | 1.51 | 0.08 |
| CK | | | 23.40 | 1.35 | 940 | 0.36 | 15.90 | 3.57 | 0.77 | 0.63 | 0.20 | | | 1.47 | 0.07 |
| CR | -0.15 | 0.03 | 24.10 | 1.36 | 656 | 0.38 | 15.80 | 2.32 | 0.30 | 1.29 | 0.07 | 0.09 | 0.05 | 1.53 | 0.08 |
| CH | -0.16 | 0.05 | | 2.57 | | | | 2.11 | 0.24 | 1.44 | 0.10 | | | | |
| CB | -0.19 | 0.02 | | 5.18 | | | | 1.77 | 0.18 | 1.23 | 0.18 | | | | |
| TL | -0.11 | 0.03 | 18.90 | 1.07 | 514 | 0.28 | 11.60 | 2.69 | 0.15 | 1.26 | 0.21 | 0.64 | 0.23 | 1.63 | 0.08 |
| TD | -0.14 | 0.02 | 20.47 | 1.16 | 587 | 0.33 | 12.66 | 2.63 | 0.07 | 1.19 | 0.07 | 0.64 | 0.13 | 1.62 | 0.08 |
| CV3 CAIs | 0.61 | 0.09 | 0.64 | 0.01 | 6042 | 0.02 | 12.56 | 8.57 | 0.40 | 5.95 | 0.51 | | | 0.05 | 0.00 |
| AOAs | 0.61 | 0.09 | 5.03 | 0.23 | 2200 | 0.20 | 19.72 | 7.77 | 0.90 | 5.12 | 0.80 | | | 0.26 | 0.01 |
| CV/CM/CO/CK chondrules | -0.08 | 0.03 | 11.00 | 0.66 | 1406 | 0.36 | 20.55 | 2.42 | 0.36 | 0.57 | 0.14 | | | 0.54 | 0.03 |
| CR chondrules | | | 14.15 | 1.04 | 1349 | 0.46 | 20.47 | 2.02 | 0.65 | 1.42 | 0.04 | | | 0.69 | 0.03 |
| FeNi metal | 0.61 | 0.09 | 93.39 | 6.31 | | | | | | | | | | | |

**Table S4. Average Ni isotopic composition of meteoritic materials.** Uncertainties are 2 standard deviations (2s.d.) for N≤3 and 95% confidence intervals (95% CI) for N ≥4, where N is the number of samples. NC and CC stand for non-carbonaceous (inner disk) and carbonaceous (outer disk) reservoir, respectively. Data is internally normalized to either $^{61}Ni/^{58}Ni$ = 0.016744 or $^{62}Ni/^{61}Ni$ = 3.1884. Data sources: (***28–34, 66, 87, 88***).

| Sample | Reservoir | N | Norm. $^{61}Ni/^{58}Ni$ | | | Norm. $^{62}Ni/^{61}Ni$ | | |
|---|---|---|---|---|---|---|---|---|
| | | | $\mu^{60}Ni$ | $\mu^{62}Ni$ | $\mu^{64}Ni$ | $\mu^{58}Ni$ | $\mu^{60}Ni$ | $\mu^{64}Ni$ |
| Hayabusa2 | | | | | | | | |
| Ryugu A | CC | 2 | -3 ± 3 | 18 ± 3 | 50 ± 4 | 55 ± 10 | 15 ± 1 | -2 ± 13 |
| Ryugu C | CC | 2 | -1 ± 5 | 26 ± 2 | 71 ± 14 | 80 ± 7 | 25 ± 2 | -6 ± 7 |
| Ryugu mean | CC | 4 | -2 ± 3 | 22 ± 8 | 60 ± 20 | 67 ± 24 | 20 ± 9 | -4 ± 8 |
| Chondrites | | | | | | | | |
| CI | CC | 6 | 0 ± 2 | 23 ± 4 | 68 ± 7 | 72 ± 11 | 24 ± 5 | -2 ± 7 |
| CM | CC | 8 | -11 ± 3 | 11 ± 1 | 35 ± 18 | 33 ± 3 | 0 ± 3 | 3 ± 16 |
| CO | CC | 4 | -7 ± 2 | 9 ± 6 | 24 ± 7 | 29 ± 20 | 2 ± 9 | -4 ± 13 |
| CV | CC | 9 | -11 ± 1 | 10 ± 4 | 27 ± 12 | 29 ± 14 | -3 ± 4 | -4 ± 3 |
| CR | CC | 4 | -15 ± 3 | 9 ± 8 | 28 ± 15 | 28 ± 24 | -6 ± 6 | 1 ± 9 |
| CH | CC | 1 | -16 ± 5 | 12 ± 11 | 28 ± 22 | 38 ± 35 | -4 ± 16 | -9 ± 17 |
| CB | CC | 1 | -19 ± 2 | 16 ± 5 | | 49 ± 14 | -3 ± 5 | |
| Tagish Lake | CC | 2 | -11 ± 3 | 15 ± 5 | 37 ± 8 | 45 ± 16 | 4 ± 8 | -5 ± 9 |
| Tarda | CC | 2 | -14 ± 2 | 16 ± 4 | 36 ± 10 | 51 ± 13 | 3 ± 6 | -6 ± 9 |
| EH | NC | 4 | -2 ± 3 | 3 ± 2 | 13 ± 13 | 8 ± 9 | 1 ± 3 | 7 ± 24 |
| EL | NC | 3 | 0 ± 7 | -3 ± 7 | -4 ± 5 | -8 ± 23 | -6 ± 10 | 3 ± 27 |
| OC | NC | 16 | -5 ± 1 | -6 ± 2 | -16 ± 4 | -19 ± 6 | -11 ± 2 | 2 ± 3 |
| R | NC | 1 | -7 ± 2 | -5 ± 3 | -8 ± 10 | -16 ± 8 | -13 ± 3 | 7 ± 11 |
| Achondrites/ Iron meteorites | | | | | | | | |
| HED | NC | 9 | -10 ± 13 | 3 ± 10 | | 11 ± 27 | -7 ± 15 | |
| Ureilites | NC | 2 | 2 ± 8 | -5 ± 16 | | -14 ± 48 | -3 ± 7 | |
| Angrites | NC | 15 | -2 ± 9 | 2 ± 4 | | 5 ± 10 | 4 ± 13 | |
| Aubrites | NC | 1 | 8 ± 8 | 5 ± 19 | | 16 ± 52 | 13 ± 19 | |
| Main group pallasites | NC | 1 | -2 ± 6 | -6 ± 10 | -13 ± 19 | -20 ± 27 | -8 ± 10 | 6 ± 30 |
| NWA 5363/5400 | NC | 1 | 0 ± 2 | 1 ± 3 | 4 ± 8 | 3 ± 8 | 1 ± 3 | 1 ± 10 |
| IAB | NC | 6 | -3 ± 3 | -5 ± 6 | -8 ± 7 | -16 ± 19 | -9 ± 9 | 7 ± 12 |
| IC | NC | 7 | -5 ± 1 | -5 ± 5 | -21 ± 13 | -17 ± 16 | -10 ± 6 | -5 ± 4 |
| IIAB | NC | 2 | -4 ± 1 | -9 ± 9 | -29 ± 24 | -29 ± 28 | -13 ± 8 | -2 ± 3 |
| IIIAB | NC | 3 | -6 ± 3 | -11 ± 8 | -35 ± 8 | -42 ± 3 | -20 ± 4 | 1 ± 6 |
| IIIE | NC | 5 | -6 ± 2 | -7 ± 3 | -25 ± 13 | -22 ± 11 | -14 ± 4 | -5 ± 5 |
| IVA | NC | 5 | -6 ± 2 | -9 ± 3 | -29 ± 14 | -27 ± 9 | -15 ± 2 | -3 ± 6 |
| IIC | CC | 4 | -16 ± 4 | 15 ± 11 | 36 ± 33 | 47 ± 35 | -1 ± 14 | -9 ± 10 |
| IID | CC | 10 | -11 ± 2 | 9 ± 8 | 23 ± 14 | 29 ± 25 | -1 ± 10 | -5 ± 15 |
| IIF | CC | 1 | -9 ± 1 | 9 ± 4 | 17 ± 11 | 28 ± 11 | 0 ± 4 | -10 ± 14 |
| IIIF | CC | 3 | -11 ± 4 | 12 ± 7 | 25 ± 18 | 36 ± 22 | 1 ± 3 | -10 ± 3 |
| IVB | CC | 21 | -13 ± 1 | 7 ± 3 | 24 ± 5 | 23 ± 8 | -6 ± 4 | 2 ± 5 |
| Wiley (IIC/ungr.) | CC | 1 | -17 ± 3 | 13 ± 5 | 34 ± 10 | 40 ± 14 | -4 ± 5 | -4 ± 15 |
| Earth's mantle | | 2 | -1 ± 1 | 4 ± 1 | 12 ± 2 | 11 ± 3 | 3 ± 1 | 1 ± 3 |
| Mars' mantle | | 5 | -1 ± 1 | 4 ± 3 | | 12 ± 9 | 3 ± 3 | |

**Table S5. Fe and Ni isotopic data for selected CAIs.** Uncertainties represent those reported in the literature and 95% confidence intervals (95% CI) for the calculated CAI mean. Data sources: **(*29, 68, 81, 89*)**.

| Sample | Host chondrite | | Norm. $^{57}Fe/^{56}Fe$ | | | Norm. $^{61}Ni/^{58}Ni$ | | | Norm. $^{62}Ni/^{61}Ni$ | | |
|---|---|---|---|---|---|---|---|---|---|---|---|
| | | $\delta^{56}Fe$ | $\mu^{54}Fe$ | $\mu^{58}Fe$ | $\delta^{60}Ni$ | $\mu^{60}Ni$ | $\mu^{62}Ni$ | $\mu^{64}Ni$ | $\mu^{58}Ni$ | $\mu^{60}Ni$ | $\mu^{64}Ni$ |
| Egg-2 | $CV3_{ox}$ Allende | 9.8 ± 1.1 | -560 ± 450 | 3400 ± 1200 | 4.15 ± 0.06 | 53 ± 6 | 143 ± 11 | 336 ± 21 | 443 ± 30 | 199 ± 11 | -86 ± 33 |
| Egg 2 high-Fe Px | $CV3_{ox}$ Allende | 9.86 ± 0.04 | -581 ± 58 | 324 ± 254 | 4.24 ± 0.06 | 60 ± 8 | 169 ± 18 | 389 ± 60 | 525 ± 49 | 232 ± 18 | -110 ± 68 |
| 31E-1 | $CV3_{red}$ Efremovka | | | | 1.33 ± 0.02 | 70 ± 2 | 159 ± 6 | 400 ± 13 | 493 ± 15 | 232 ± 5 | -70 ± 18 |
| 31E-2 | $CV3_{red}$ Efremovka | | | | 1.77 ± 0.03 | 68 ± 7 | 161 ± 12 | 428 ± 11 | 500 ± 33 | 232 ± 12 | -49 ± 32 |
| 31E-4 | $CV3_{red}$ Efremovka | | | | 1.27 ± 0.04 | 56 ± 2 | 130 ± 5 | 349 ± 9 | 402 ± 14 | 189 ± 5 | -35 ± 15 |
| CAI mean | | | -581 ± 56 | | | 61 ± 9 | 152 ± 20 | 380 ± 47 | 473 ± 61 | 217 ± 27 | -70 ± 37 |